\documentclass{Interspeech}
\usepackage{cite}
\usepackage{amsmath,amssymb,amsfonts}
\usepackage{algorithmic}
\usepackage{graphicx}
\usepackage{textcomp}
\usepackage{xcolor}
\usepackage{setspace}
\usepackage{subcaption}
\usepackage{enumerate}
\usepackage{hyperref}
\usepackage{booktabs}
\usepackage{multirow}
\usepackage{float}
\usepackage{hyperref}
\usepackage{enumitem}

\interspeechcameraready

\title{Multi-Channel Acoustic Echo Cancellation Based on Direction-of-Arrival Estimation}

\author[affiliation={1,2}]{Fei}{Zhao}
\author[affiliation={1}]{Xueliang}{Zhang}
\author[affiliation={2}]{Zhong-Qiu}{Wang}

\affiliation{College of Computer Science}{Inner Mongolia University}{China}
\affiliation{Department of Computer Science and Engineering}{Southern University of Science and Technology}{China}
\email{zhaofei@mail.imu.edu.cn, cszxl@imu.edu.cn, wang.zhongqiu41@gmail.com}

\usepackage{comment}

\begin{document}

\maketitle

\begingroup
\renewcommand\thefootnote{}\footnote{This work was done while Fei Zhao was a visiting student at SUSTech. \textit{Corresponding author: Zhong-Qiu Wang}.}
\addtocounter{footnote}{0}
\endgroup

\begin{abstract}
Acoustic echo cancellation (AEC) is an important speech signal processing technology that can remove echoes from microphone signals to enable natural-sounding full-duplex speech communication.
While single-channel AEC is widely adopted, multi-channel AEC can leverage spatial cues afforded by multiple microphones to achieve better performance.
Existing multi-channel AEC approaches typically combine beamforming with deep neural networks (DNN).
This work proposes a two-stage algorithm that enhances multi-channel AEC by incorporating sound source directional cues.
Specifically, a lightweight DNN is first trained to predict the sound source directions, and then the predicted directional information, multi-channel microphone signals, and single-channel far-end signal are jointly fed into an AEC network to estimate the near-end signal.
Evaluation results show that the proposed algorithm outperforms baseline approaches and exhibits robust generalization across diverse acoustic environments.
\end{abstract}

\section{Introduction}

In hands-free devices such as mobile phones, smart-home appliances, and teleconferencing systems, the coupling between the loudspeakers and the microphones causes the device's microphones to capture the sound played by its loudspeakers, resulting in sound echoes, which can severely degrade speech communication \cite{benesty2001advances, hansler2005acoustic}.
To address this issue, acoustic echo cancellation (AEC) is designed to estimate the acoustic echo paths from the loudspeakers to the microphones \cite{4648922, 10096597, Zhao2024}, and if the echo paths can be accurately estimated, the echoes can be reduced from the recorded microphone signals \cite{benesty2004adaptive, enzner2014acoustic}.
In the past decades, numerous AEC techniques have been developed, including adaptive filters in the time domain such as normalized least mean squares (NLMS) and recursive least squares (RLS) \cite{haykin2005adaptive}, and frequency-domain methods such as frequency-domain adaptive filters \cite{bershad1979analysis} and partitioned-block-based frequency-domain Kalman filters \cite{kuech2014state}.
Linear filtering approaches cause little distortion to near-end speech signals and they usually exhibit robust generalization across diverse acoustic conditions.
However, a significant limitation arises when dealing with nonlinear echoes, as linear methods are not very effective at canceling out nonlinear echoes, resulting in substantial residual echoes \cite{guerin2004nonlinear, halimeh2019neural, patel2024nonlinear, yin2024nonlinear}.
This limitation severely compromises communication quality, particularly in systems with low-quality amplifiers and loudspeakers.

In recent years, deep learning has been increasingly applied to AEC in an end-to-end (E2E) manner, demonstrating remarkable effectiveness and strong potential \cite{franzen2021aec, zhang2022complex, 10096411, 10141581}.
These methods frame AEC as a speech separation or enhancement task, training deep neural networks (DNN) to directly extract near-end speech from microphone signals containing echoes.
However, over-suppression of echo components may result in excessive cancellation of target near-end speech, introducing distortions that hinder the practical utility of DNN-based AEC models for downstream applications \cite{9413585, 9746272, Panchapagesan2022}.
In addition, existing DNN-based approaches exhibit limited generalization across diverse acoustic conditions, posing significant challenges for real-world deployment \cite{sridhar2021icassp, saka2023conversational, Wang2022GridNetJournal}.

With the ongoing advances in speech signal processing technologies, single-channel systems encounter many challenges when processing speech signals in complex acoustic conditions.
Multi-channel systems, by leveraging multiple microphones to exploit the spatial information of different sound sources, can offer richer acoustic information, thereby demonstrating substantial theoretical advantages and capable of producing superior performance in complex acoustic environments \cite{benesty2008microphone}.
In multi-channel scenarios, traditional AEC algorithms face non-unique solution problems due to the strong correlation among multiple echo signals \cite{sondhi1995stereophonic, benesty1998better}, rendering the multi-channel adaptive filter ill-posed and unstable \cite{helwani2013multichannel}, while DNN-based AEC followed by beamforming can alleviate the above problems and improve the performance\cite{zhang2022multi, zhang2023neural}.

In this paper, we propose a novel multi-channel AEC algorithm, which first predicts the direction of the sound source through a DNN and then utilizes the predicted directional information to enhance AEC performance.
Our method consists of two stages.
In the first stage, a lightweight DNN is trained to predict the directions of the sound sources, including both the loudspeakers and the near-end target talker.
To align with the real-time requirements of AEC, direction prediction is performed at the frame level in a causal, frame-online manner.
In the second stage, the predicted directional information, multi-channel microphone signals, and single-channel far-end signal are combined and fed into an AEC network to estimate the near-end target signal.
We investigate various methods for integrating directional information and analyze their benefits and limitations. 
Evaluation results show that the proposed method outperforms several representative baselines and exhibits robust generalization across multiple mismatched test sets.


\section{Problem Formulation}

In a teleconferencing system with $P$ near-end loudspeakers and $Q$ microphones, the far-end time-domain signal $x$ is played through the loudspeakers.
Let us denote the signal played out by loudspeaker $p\in \{1, 2, \dots, P\}$ as $x^\text{nl}_p$, where the superscript ``$\text{nl}$'' means that the far-end signal is nonlinearly distorted by the loudspeaker.
The mixture signal captured by microphone $q$ (i.e., $y_q$ with $q\in\{1,2,\dots, Q\}$ indexing the $Q$ microphones) can be modeled, in the time domain, as
\begin{align}
y_q &= \sum\nolimits_{p=1}^{P} h_{pq} * x^\text{{nl}}_p + s_q \\
&= \sum\nolimits_{p=1}^{P} e_{pq} + s_q \\
&= \sum\nolimits_{p=1}^{P} e_{pq} + s_q^{\text{d}} + s_q^{\text{r}}.
\end{align}

In the second row, the echo signal $e_{pq}=h_{pq} * x^\text{{nl}}_p$ represents the signal image from loudspeaker $p$ to microphone $q$, with $h_{pq}$ denoting the corresponding echo path and $*$ the convolution operation.
In the third row, the reverberant near-end speech $s_q=s^{\text{d}}_q+s^{\text{r}}_q$ consists of direct component $s^{\text{d}}_q$ and reverberant component $s_q^{\text{r}}$. 

Without loss of generality, we designate microphone $1$ as the reference microphone.
Our goal is to recover $s_1^{\text{d}}$ based on the multi-channel mixture signal $\{y_q\}^Q_{q=1} $ and the far-end signal $x$.
The signal played out by the loudspeakers significantly impacts the quality of the recorded mixtures, and the nonlinear distortion introduced by the loudspeakers poses significant challenges for accurate AEC.

\vspace{-0.1cm}
\section{Method}

\begin{figure}
        \centering
        \vspace{-0.3cm}
	\includegraphics[width=0.92\linewidth]{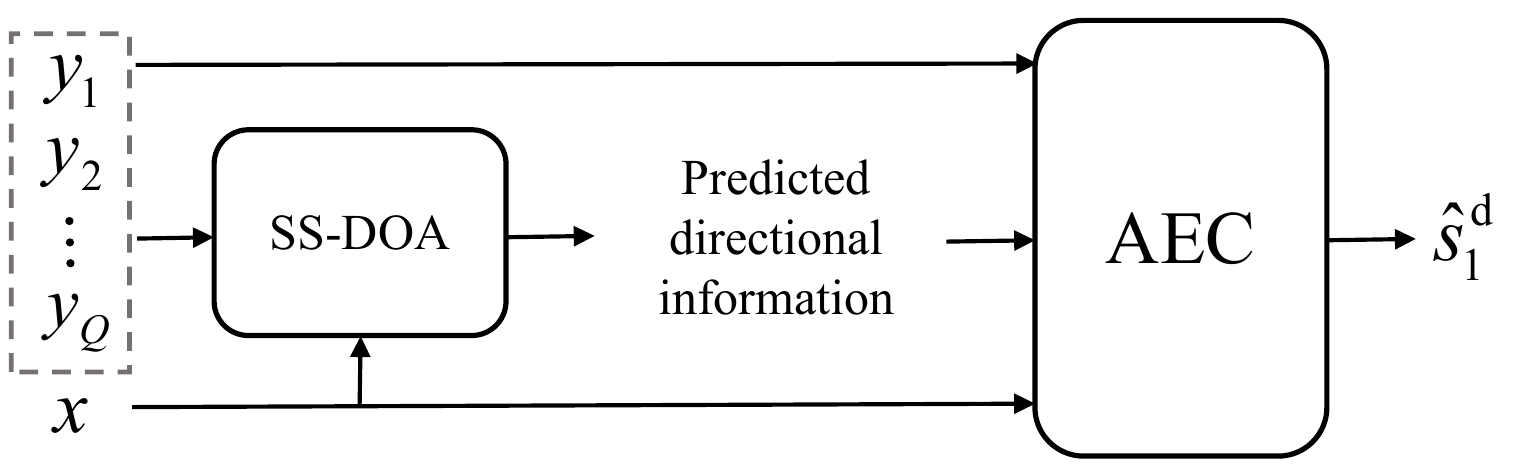}
	\caption{Overview of proposed method.}
        \vspace{-0.55cm}
	\label{fig:overview}
\end{figure}

The proposed system is illustrated in \autoref{fig:overview}, which consists of two modules, one for sound source direction of arrival estimation (SS-DOA) and the other for AEC.

\vspace{-0.1cm}
\subsection{SS-DOA module}
\label{subsec: ss-doa}


The SS-DOA module predicts the DOA of the two loudspeakers and the target talker at the frame level in a causal, frame-online fashion.
The system is trained to predict $36$ hypothesized source directions, which are evenly distributed between $0$ and $360$ degrees with an interval of $10$ degrees.

\begin{figure}
        \centering
        \vspace{-0.3cm}
	\includegraphics[width=0.92\linewidth]{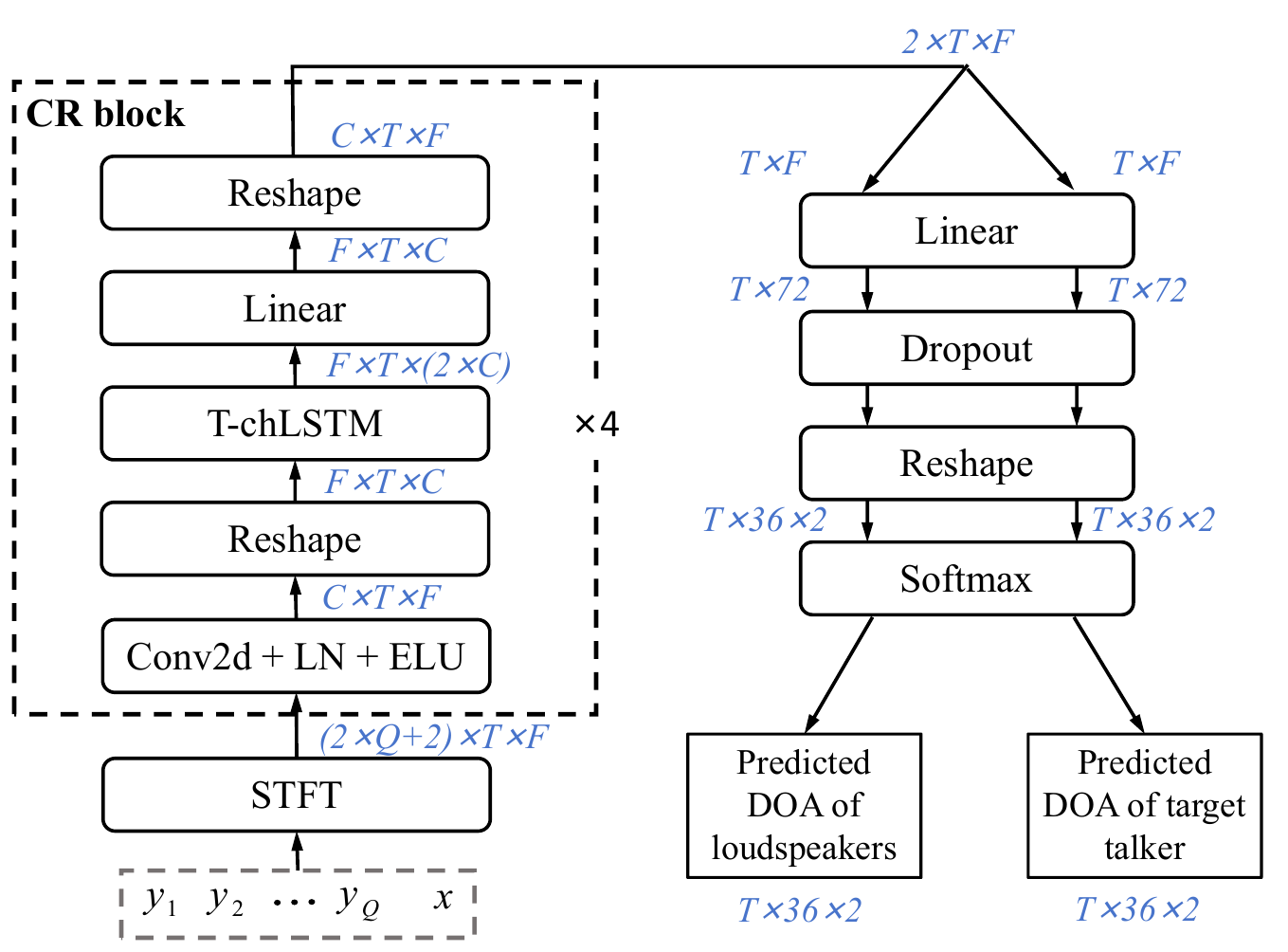}
	\caption{Illustration of SS-DOA module.}
        \vspace{-0.55cm}
	\label{fig:DOA}
\end{figure}

\autoref{fig:DOA} shows the DNN architecture of SS-DOA.
The network stacks the real and imaginary (RI) components of the $Q$-microphone input signals and the far-end signal as input, which is passed through $4$ convolutional-recurrent (CR) blocks.
The CR block is first built by a causal 2D convolution (Conv2D) layer with a kernel size of $3 \times 3$ to obtain a $C$-dimensional embedding for each T-F unit, resulting in a $C \times T \times F$ tensor.
For the first CR block, the Conv2D layer transforms the input embedding from $(2\times Q+2)\text{-}$ to $C\text{-}$dimensional,
which is further processed by a layer normalization (LN) layer and an exponential linear unit (ELU) activation function.
Next, we consider the resulting $C\times T\times F$ tensor as $F$ sequences, each with length $T$ and a $C\text{-}$dimensional embedding at each time step, and we use a sub-band time dimension channel LSTM (T-chLSTM) \cite{DBLP:conf/icassp/LiuZ23d}, which has $2\times C$ hidden units and is shared by all the $F$ frequencies, to model the $C\times T$ tensor at each frequency.
After that, for the CR blocks that are not the last one, a linear layer is applied to map the $(2\times C)$-dimensional embedding to $C\text{-}$dimensional, and the tensor is reshaped back to $C \times T \times F$.
For the last CR block, the Conv2D layer compresses the channel dimension from $C$ to $2$, and the linear layer compresses the embedding dimension from $2\times C$ to $2$,
obtaining a $2\times T\times F$ tensor for the subsequent classification-based DOA estimation block.
In the DOA estimation block, we first split the $2\times T\times F$ tensor into two $T\times F$ tensors, which are respectively utilized for estimating the DOA of the loudspeakers and that of the target talker.
Next, for each of the two splited tensors, the $F$-dimensional embedding at each frame is linearly mapped to $72$-dimensional, followed by a dropout operation with a dropout ratio of $0.2$ and then a reshape operation converting the last dimension from $72$ to $36 \times 2$, which is designed to indicate whether a sound source exists for each one of the $36$ hypothesized directions.

We set two $36$-directional binary classifiers to respectively predict the DOA of the loudspeakers and that of the target talker at the frame level.
The classification results indicate whether a sound source exists for each of the $36$ directions.
Specifically, at each direction, a predicted tensor of $[0,1]$ (obtained after thresholding) indicates the absence of a sound source in the current direction, while $[1,0]$ signifies its presence.
If all the $36$ directions are classified as absent of a sound source, it indicates that this frame is silent.
Considering that, in reality, two loudspeakers rarely co-locate in the same direction, predicting two distinct directions among the $36$ hypothesized directions identifies the sound source directions of the two loudspeakers.

\subsection{AEC module}\label{AEC_module_description}

\begin{figure}
        \centering
        \vspace{-0.3cm}
	\includegraphics[width=0.92\linewidth]{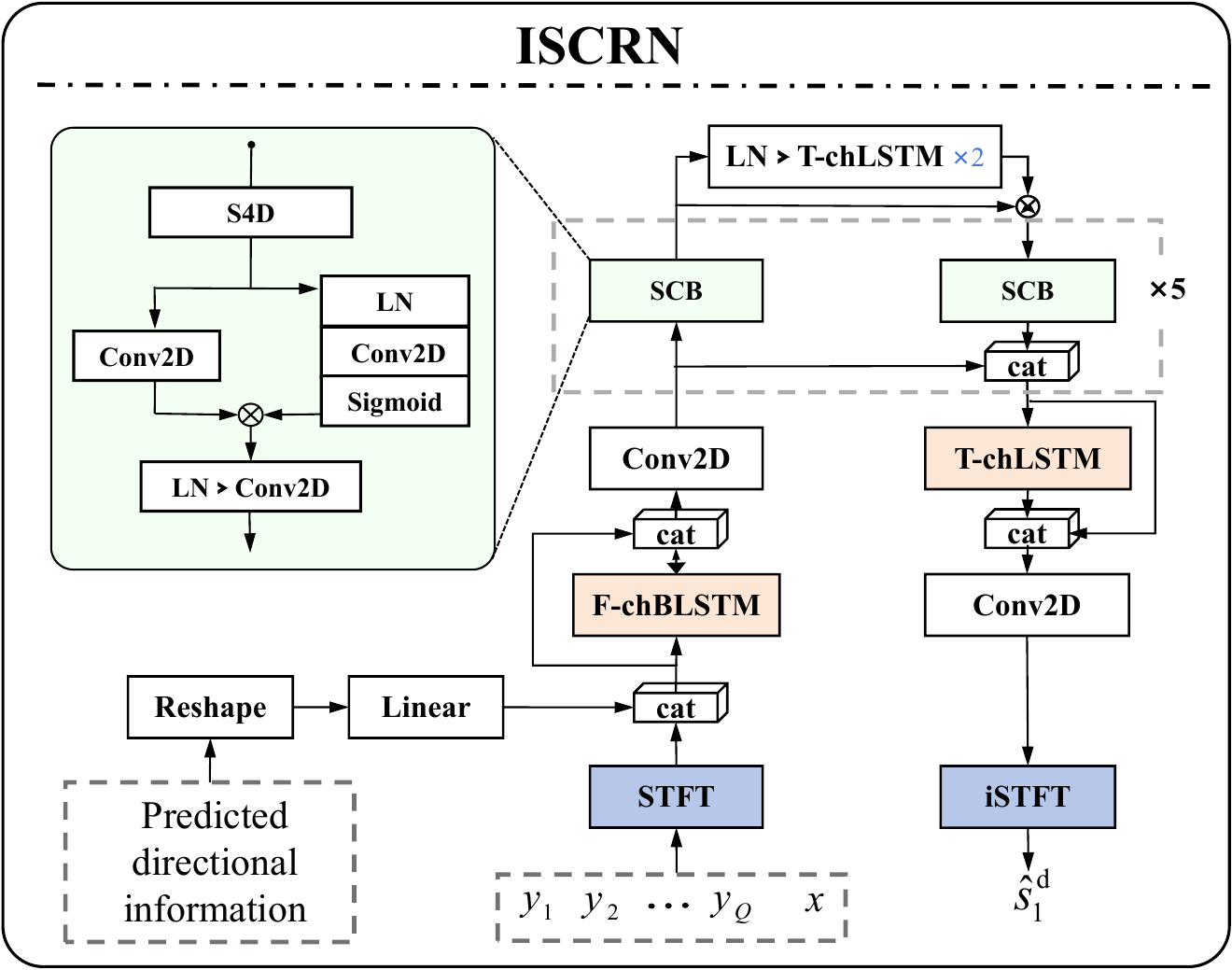}
	\caption{Illustration of AEC module, which can leverage predicted directional information produced by SS-DOA module to improve AEC.}
        \vspace{-0.645cm}
	\label{fig:AEC}
\end{figure}

The AEC module is illustrated in \autoref{fig:AEC}.
It uses a structured state-space module (S4D) to replace the cepstrum unit in ICCRN \cite{DBLP:conf/icassp/LiuZ23d}, resulting in ISCRN.
As an in-place model, ISCRN does not involve frequency down- or up-sampling.
This reduces information loss caused by down-sampling and avoids the introduction of irrelevant information by up-sampling.\footnote{The details of ISCRN can be found at \url{https://github.com/ZhaoF-i/DOA-AEC}.}

In this study, we explore three different modes of incorporating predicted directional information to improve AEC:
\begin{description}
[noitemsep, topsep=0pt]
\item[a.] The output embedding from the last CR block in \autoref{fig:DOA} (i.e., the $2\times T\times F$ tensor) serves as the predicted directional information.
This embedding is concatenated with the multi-channel microphone signal and the single-channel far-end signal in the time-frequency domain, and the combined input is fed into the AEC network to predict the target signal.
See \autoref{fig:AEC} for an illustration.
The embedding encodes the directional information of both the loudspeakers and the target talker, enabling the exploration of whether richer directional cues can enhance the subsequent AEC performance.
\item[b.] Compared to mode \textbf{a}, only the embedding representing the target talker (i.e., the right $T\times F$ tensor separated from the $2\times T\times F$ tensor in \autoref{fig:DOA}) is used as direction information to investigate the impact of explicit target direction cues.
\item[c.] The final DOA estimate of the target talker branch in \autoref{fig:DOA} is used as the directional information. The tensor, which has a shape of $T\times 36\times 2$, is reshaped to $T \times 72$, linearly mapped to $T \times F$, and merged with the speech signal (as shown in \autoref{fig:AEC}). Compared to \textbf{b}, this approach includes an additional softmax activation function, resulting in a clearer representation of direction information.
\end{description}

\vspace{-0.1cm}
\subsection{Loss function}

For the SS-DOA module, following \cite{cooreman2023crnn, nie2024multi} we design a binary cross-entropy (BCE) loss, which contrasts true class labels with predicted probabilities:
\begin{align}
\mathcal{L}_{\text{DOA}} = \texttt{BCE}(A_{s}, \hat{A}_{s}) + \texttt{BCE}(A_{t}, \hat{A}_{t}).
\end{align}
where $A_s$ and $\hat{A}_s$ respectively denote the real and predicted loudspeaker directions, and $A_t$ and $\hat{A}_t$ the real and predicted near-end talker directions.

For the AEC module, the model is trained by using a magnitude loss and an RI loss \cite{10447755, zhao2024attention}:
\begin{equation}
	\mathcal{L}_{\text{RI+Mag}} = \mathcal{L}_{\mathrm{RI}} + \mathcal{L}_{\text{Mag}},
\end{equation}
\begin{equation}
	\mathcal{L}_{\mathrm{RI}} = \sum_{t,f} \Big| |S^{\text{d}}_1(t, f)|^p e^{j \angle{S^{\text{d}}_1(t, f)}}-|\hat{S}^{\text{d}}_1(t, f)|^p e^{j \angle{\hat{S}^{\text{d}}_1(t, f)}} \Big|^2,
\end{equation}
\begin{equation}
	\mathcal{L}_{\mathrm{Mag}}=\sum_{t,f} \Big||S^{\text{d}}_1(t, f)|^p-|\hat{S}^{\text{d}}_1(t, f)|^p\Big|^2 ,
\end{equation}
where $p$ (tuned to $0.5$) is a spectral compression factor \cite{li2021importance}.

\section{Experimental Setup}
\label{sec:EXPERIMENT}

\begin{table*}
  \centering
  \caption{Comparison of PESQ, SDR (dB) and ERLE (dB) scores on unmatched test sets (best scores are highlighted in bold).}\vspace{-0.3cm}
  \scalebox{1.0}{
\fontsize{6.5}{8}\selectfont
    \begin{tabular}{lccccccccccccccc}
    \toprule
    Testset type & \multicolumn{5}{c}{Near-end Talker Position Change}               & \multicolumn{5}{c}{1° Sound Source Angle Interval}                   & \multicolumn{5}{c}{Co-directional Near-end Talker and Loudspeaker} \\
    \cmidrule(lr){2-6}\cmidrule(lr){7-11}\cmidrule(lr){12-16}
    Test scenarios & \multicolumn{2}{c}{DT} & \multicolumn{2}{c}{ST\_NE} & \multicolumn{1}{l}{ST\_FE} & \multicolumn{2}{c}{DT} & \multicolumn{2}{c}{ST\_NE} & \multicolumn{1}{l}{ST\_FE} & \multicolumn{2}{c}{DT} & \multicolumn{2}{c}{ST\_NE} & \multicolumn{1}{l}{ST\_FE} \\
    \cmidrule(lr){2-3}\cmidrule(lr){4-5}\cmidrule(lr){6-6}
    \cmidrule(lr){7-8}\cmidrule(lr){9-10}\cmidrule(lr){11-11}\cmidrule(lr){12-13}\cmidrule(lr){14-15}\cmidrule(lr){16-16}
    Model & PESQ    & SDR     & PESQ    & SDR     & ERLE    & PESQ    & SDR     & PESQ    & SDR     & ERLE    & PESQ    & SDR     & PESQ    & SDR     & ERLE \\
    \midrule
    Mixture     & 1.55    & -10.6  & 2.24    & -5.2   & --      & 1.54    & -11.2  & 2.28    & -5.0   & --      & 1.54    & -11.1  & 2.29    & -5.1    & -- \\
    \midrule
    ISCRN & 2.50     & 4.2    & 2.92    & 4.7    & 60.77   & 2.50     & 4.3    & 2.96    & 5.5    & 60.79   & 2.50     & 4.2    & 2.97    & 5.1    & 60.73 \\
    ISCRN + DI$_\text{B}$ & 2.60     & 4.3    & 2.99    & 4.8    & 59.75   & 2.58    & 4.5    & 3.03    & 5.8    & 59.82   & 2.60     & 4.6    & 3.06    & 5.7    & 59.66 \\
    \midrule
    ISCRN + DI$_\text{E}$ & 2.53    & 4.0    & 2.90     & 4.5     & \textbf{61.24}   & 2.60     & 4.4    & 3.01    & 5.5     & \textbf{61.33}   & 2.55    & 4.3    & 3.02    & 5.1    & \textbf{61.18} \\
    ISCRN + DI$_\text{ET}$ & \textbf{2.67}    & 4.4     & \textbf{3.03}    & 4.8    & 59.93   & \textbf{2.74}    & 4.9    & \textbf{3.12}    & 5.9    & 60.11   & \textbf{2.60}    & 4.8    & \textbf{3.13}    & 5.5    & 59.97 \\
    ISCRN + DI$_\text{ETA}$ & 2.64    & \textbf{4.7}    & 2.97    & \textbf{5.3}    & 60.84   & 2.69    & \textbf{5.5}    & 3.10     & \textbf{6.7}    & 60.87   & 2.65    & \textbf{5.3}    & 3.12    & \textbf{6.4}    & 60.70 \\
    \bottomrule
    \end{tabular}%
    }
  \label{tab:unmatch_test}%
  \vspace{-0.37CM}
\end{table*}%

\subsection{Datasets}

\begin{figure}
        \centering
        \vspace{-0.3cm}
	\includegraphics[width=0.92\linewidth]{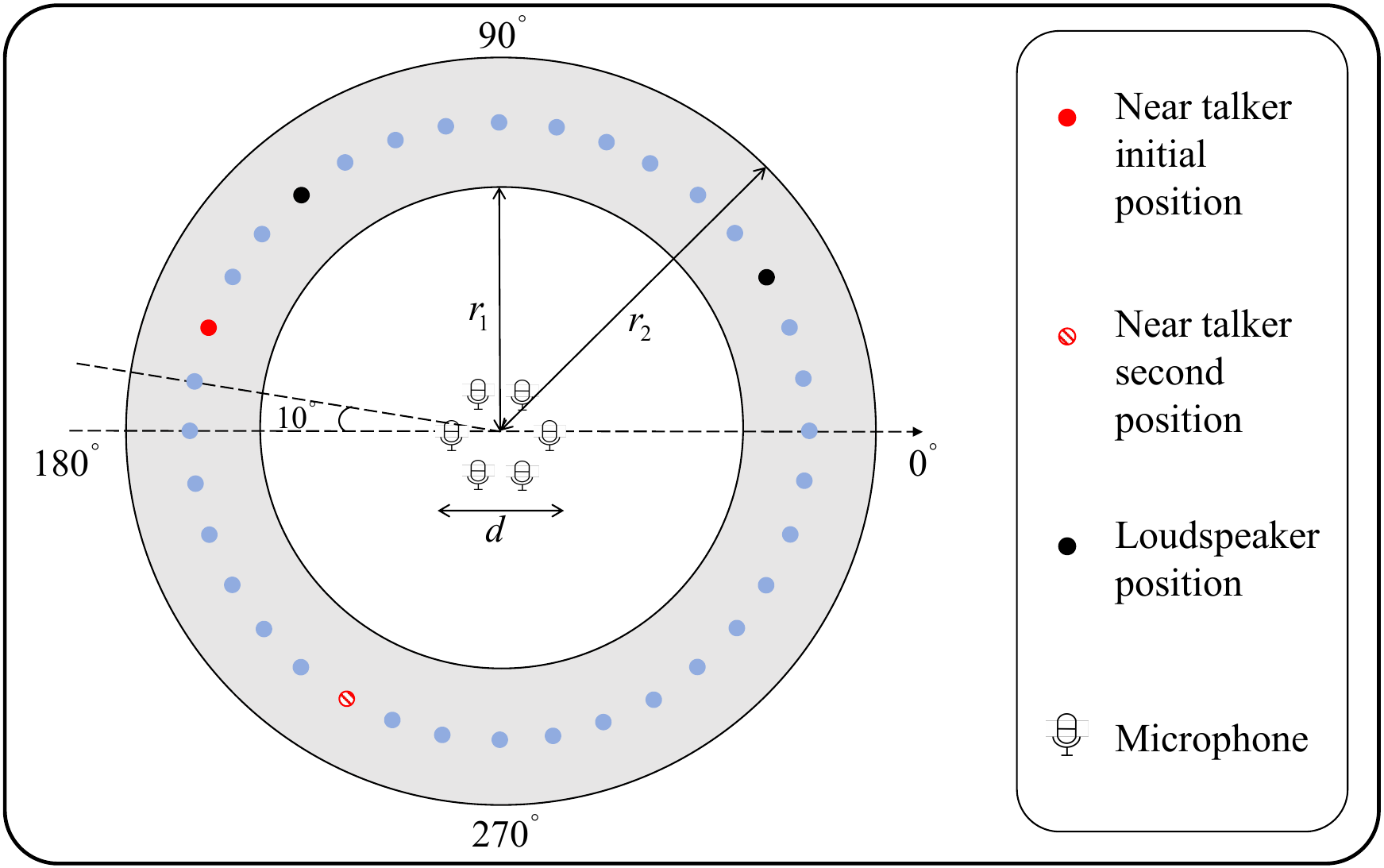}
	\caption{
    Scenario illustration. Best viewed in color.
}
        \vspace{-0.56cm}
	\label{fig:dataset}
\end{figure}

The near-end and far-end signals employed in our experiments are sourced from the synthetic datasets of the ICASSP $2023$ AEC challenge \cite{DBLP:journals/corr/abs-2309-12553} to simulate the near-end and far-end scenarios.
The room impulse responses (RIR) are generated by using the image method \cite{allen1979image}.
We simulate different rooms of dimensions $l \times w \times h$, with $l$ randomly-sampled from $4$ to $8$ meters, $w$ from $3$ to $7$ meters, and $h$ from $3$ to $5$ meters.
\autoref{fig:dataset} illustrates the relationship between the microphone array and the sound source positions.
The array consists of $6$ microphones ($Q=6$) uniformly arranged on a circle with a diameter of $d=7$ cm, and it is placed at the room center.
Two loudspeakers ($P=2$) and a near-end target talker are positioned at the same height as the array.
Their directions are randomly selected within a $360$-degree range (in $10$-degree intervals) without repetition.
The distance from each sound source to the array center is randomly sampled from the range $[r_1, r_2]$ meters, where $r_1$ is fixed at $1$ meter and $r_2$ does not exceed the room dimensions.
The reverberation time (T$60$) is randomly sampled from $0.1$ to $0.8$ seconds for simulating rooms to cover a wide range of acoustic conditions.
In total, $22,000$ RIRs are generated based on the aforementioned configuration, and $20,000$ of them are allocated for training, $1,000$ for validation, and $1,000$ for testing.
The generated echo signal is mixed with near-end speech $s$ at a signal-to-echo ratio (SER) randomly sampled from the range $[-10, 10]$ dB with a step of $1$ dB.
The nonlinearity setting adheres to the configurations established in \cite{DBLP:conf/icassp/ZhangLZ22}.
We generate a dataset consisting of $100,000$ training samples, $5,000$ validation samples, and $1,000$ test samples via a random selection process.
All speech samples have a duration of $6$ seconds.
The sampling rate is $16$ kHz. 

To evaluate the generalizability of the proposed method, we construct three unmatched test sets.
The first one, as illustrated in \autoref{fig:dataset}, changes the position of the near-end target talker while the talker is speaking, where the talker is fixed at the initial position in the first $3$ seconds while moved to the second position afterward.
The second one reduces the angle interval size of the original test set from $10$ degrees to $1$ degree, ensuring that different sound sources do not overlap. Reducing the angular interval can increase the resolution of the test set to verify the ability of the algorithm on the high-resolution test set.
The third one puts the near-end target talker in the same direction as one of the loudspeakers. This setup simulates scenarios where the target sound source aligns with the direction of background noise (e.g., another loudspeaker), enabling the evaluation of the algorithm's performance under strong interference conditions.
Each of these test sets contains $1,000$ samples.

\begin{table}[]
  \centering
  \caption{Comparison of PESQ, SDR (dB), and ERLE (dB) scores on the matched test set (best scores are bolded).}\vspace{-0.2cm}
  \scalebox{1.0}{
\fontsize{6.8}{8}\selectfont
    \begin{tabular}{lccccc}
    \toprule
    Test scenarios & \multicolumn{2}{c}{DT}         & \multicolumn{2}{c}{ST\_NE}        & \multicolumn{1}{l}{ST\_FE} \\
    \cmidrule(lr){2-3}\cmidrule(lr){4-5}\cmidrule(lr){6-6}
    Model & \multicolumn{1}{l}{PESQ       } & \multicolumn{1}{l}{SDR       } & \multicolumn{1}{l}{PESQ       } & \multicolumn{1}{l}{SDR       } & \multicolumn{1}{l}{ERLE   } \\
    \midrule
    Mixture     & 1.57    & -10.7  & 2.26    & -5.5   & \multicolumn{1}{c}{--} \\
    \midrule
    ISCRN & 2.53    & 4.5    & 2.96    & 5.3    & 60.77 \\
    ISCRN + DI$_\text{B}$ & 2.64    & 4.8    & 3.05    & 5.9     & 59.75 \\
    \midrule
    ISCRN + DI$_\text{E}$ & 2.64    & 4.7     & 3.02    & 5.4    & \textbf{61.24} \\
    ISCRN + DI$_\text{ET}$ & \textbf{2.76}    & 5.0    & \textbf{3.14}    & 5.8    & 59.93 \\
    ISCRN + DI$_\text{ETA}$ & 2.75    & \textbf{5.7}    & 3.12    & \textbf{6.8}    & 60.84 \\
    \bottomrule
    \end{tabular}%
    }
  \label{tab:test}%
  \vspace{-0.2cm}
\end{table}%

\begin{table}
  \centering
  \caption{Classification performance of SS-DOA.} \vspace{-0.2cm}
  \scalebox{0.85}{
\fontsize{6.8}{8}
    \begin{tabular}{cccc}
    \toprule
    Sound sources & Precision & Recall & F1 \\
    \midrule
    Loudspeakers & 0.751 & 0.601 & 0.636 \\
    Target talker & 0.734 & 0.601 & 0.640 \\
    \bottomrule
    \end{tabular}%
    }
  \label{tab:classification}%
  \vspace{-0.4cm}
\end{table}%

\vspace{-0.1cm}
\subsection{Training details}

For the short-time Fourier transform (STFT),
the window size is $20$ ms, the hop size $10$ ms, and the Hamming window is used as the analysis window. Set $C=20$ in Section \ref{subsec: ss-doa}.
The model is trained using Adam Optimizer \cite{DBLP:journals/corr/KingmaB14}, starting from an initial learning rate of $0.001$, which is halved if the validation loss is not decreased for two epochs.
Early stopping is applied if no improvement is observed for $10$ epochs.
No re-training or fine-tuning is performed for the three unmatched test sets.
This design can help us evaluate the generalizability of the algorithm.


\subsection{Evaluation setup}
\label{sec:Evaluation setup}

We consider three evaluation scenarios, including double talk (DT), near-end single talk (ST\_NE), and far-end single talk (ST\_FE).
Three popular metrics are used, including echo return loss enhancement (ERLE) \cite{enzner2014acoustic} for measuring echo suppression during single talk, perceptual evaluation of speech quality (PESQ) \cite{DBLP:conf/icassp/RixBHH01} for assessing near-end speech quality, and signal-to-distortion ratio (SDR) \cite{vincent2006performance} for quantifying near-end speech fidelity during double talk.

The evaluated models include ISCRN and ISCRN+DI$_*$, where DI$_*$ denotes the utilization of different directional information (DI) as additional input.
Specifically, 
\begin{itemize}[leftmargin=*,noitemsep,topsep=0pt]
\item \textbf{ISCRN}, similar to existing DNN-based AEC systems, takes the multi-channel mixture signals and the far-end signal as inputs to directly predict the near-end target signal.
\item \textbf{ISCRN+DI$_\text{B}$} employs MVDR beamforming to leverage directional information.
For the MVDR beamformer, the steering vectors are calculated using the ground truth directions, and the mixture covariance matrix is calculated in an online, streaming manner based on the mixture signals.
\item \textbf{ISCRN+DI$_\text{E}$} sends the embedding output by the last CR block (Figure 2) as directional information to the AEC model. This corresponds to the mode \textbf{a} described in Section \ref{AEC_module_description}.

\item In \textbf{ISCRN+DI$_\text{ET}$}, only the talker's directional embedding is utilized in the AEC model, corresponding to mode \textbf{b} as described in Section \ref{AEC_module_description}.

\item \textbf{ISCRN+DI$_\text{ETA}$} converts the explicitly predicted DOA of the target talker (see \autoref{fig:DOA}) into an embedding (using the method illustrated in \autoref{fig:AEC}) and sends it to the AEC model. This corresponds to the mode \textbf{c} described in Section \ref{AEC_module_description}.
\end{itemize}
ISCRN and ISCRN+ID$_\text{B}$ are considered as the baselines of the other three models.
    
\section{Evaluation Results}
\label{sec:RESULTS}

Table \ref{tab:test} and \ref{tab:unmatch_test} respectively report the performance on the matched test set and on the three unmatched test sets.

The comparison between ISCRN and ISCRN+ID$_*$ indicates that the introduction of directional information can significantly improve AEC performance. In particular,  ISCRN+ID$_\text{ET}$ has a $0.23$ improvement in PESQ compared to ISCRN in the double-talk scenario of the matching test (i.e., $2.76$ vs. $2.53$) and even has a significant improvement in the three unmatched test sets.
In addition, the analysis of ISCRN+ID$_\text{B}$ and its variants suggests that providing more explicit directional information to the network yields better results.
Notably, ISCRN+DI$_\text{E}$ obtains worse performance than ISCRN+DI$_\text{ET}$ and ISCRN+DI$_\text{ETA}$, potentially due to the network's inability to effectively utilize the additional directional information of the loudspeakers.
Compared to ISCRN+DI$_\text{ET}$, ISCRN+DI$_\text{ETA}$ incorporates a softmax function, which enhances the clarity and interpretability of the directional information.
This difference likely contributes to the observed performance variations across different evaluation metrics.
On the other hand, even without fine-tuning the trained AEC and SS-DOA modules, the proposed method outperforms the baseline models on the three unmatched test sets.

The results in Table \ref{tab:classification} demonstrate the performance of SS-DOA in distinguishing sound source directions for loudspeakers and near-end talkers, respectively. 
We utilize precision, recall, and F1 scores to report the performance of the classification-based DOA module.
While the classification performance is not strong, it still contributes to a notable improvement in the AEC performance.

It should be noted that the ISCRN network for AEC has a computational complexity of $3.64$ GMAC and a parameter size of $950.8$ thousands, and the 4CR network for DOA estimation has a computational complexity of $826.8$ MMAC and a parameter size of $92.8$ thousands.

\section{Conclusions}

We have proposed a multi-channel AEC method that utilizes directional information derived from sound source DOA estimation to improve AEC performance.
Evaluation results demonstrate that the proposed method significantly outperforms the baselines and achieves consistent performance across multiple unmatched test sets, highlighting its strong generalization capability.
Future research would focus on refining the classification network to further improve the AEC performance.


\bibliographystyle{IEEEtran}
\bibliography{mybib}

\end{document}